\documentclass[manuscript]{acmart}
\usepackage[utf8]{inputenc}

  \acmConference[]{Soft. Eng. 4 Res. Sci.}{July 23--27, 2023}{Portland,
OR}

\title{Wanted: standards for automatic reproducibility of computational
experiments}

\author{Samuel Grayson}
  
  \email{grayson5@illinois.edu}
  \orcid{0000-0001-5411-356X}
      \affiliation{%
    \institution{University of Illinois Urbana-Champaign}%
    \department{Department of Computer Science}%
    \streetaddress{201 North Goodwin Avenue MC 258}%
    \city{Urbana}%
    \state{IL}%
    \postcode{61801-2302}%
    \country{USA}%
    }

\author{Reed Milewicz}
  
  \email{rmilewi@sandia.gov}
  \orcid{0000-0002-1701-0008}
      \affiliation{%
    \institution{Sandia National Laboratories}%
    \department{Software Engineering and Research Department}%
    \streetaddress{1515 Eubank Blvd SE1515 Eubank Blvd SE}%
    \city{Albuquerque}%
    \state{NM}%
    \postcode{87123}%
    \country{USA}%
    }

\author{Joshua Teves}
  
  \email{jbteves@sandia.gov}
  \orcid{0000-0002-7767-0067}
      \affiliation{%
    \institution{Sandia National Laboratories}%
    \streetaddress{1515 Eubank Blvd SE1515 Eubank Blvd SE}%
    \city{Albuquerque}%
    \state{NM}%
    \postcode{87123}%
    \country{USA}%
    }

\author{Daniel S. Katz}
  
  \email{dskatz@illinois.edu}
  \orcid{0000-0001-5934-7525}
      \affiliation{%
    \institution{University of Illinois Urbana-Champaign}%
    \department{Department of Computer Science}\department{National
Center for Supercomputing Applications}\department{Deparment of
Electrical and Computer Engineering}\department{School of Information
Sciences}%
    \streetaddress{201 North Goodwin Avenue MC 258}%
    \city{Urbana}%
    \state{IL}%
    \postcode{61801-2302}%
    \country{USA}%
    }

\author{Darko Marinov}
  
  \email{marinov@illinois.edu}
  \orcid{0000-0001-5023-3492}
      \affiliation{%
    \institution{University of Illinois Urbana-Champaign}%
    \department{Department of Computer Science}%
    \streetaddress{201 North Goodwin Avenue MC 258}%
    \city{Urbana}%
    \state{IL}%
    \postcode{61801-2302}%
    \country{USA}%
    }

\acmYear{2023}

\keywords{reproducibility, linked data, semantic web}

\setcopyright{none}

\usepackage{hyperref}

\usepackage{color}
\usepackage{fancyvrb}

\DefineVerbatimEnvironment{Highlighting}{Verbatim}{commandchars=\\\{\}}
\newenvironment{Shaded}{}{}

\newcommand{\BuiltInTok}[1]{\textcolor[rgb]{0.00,0.50,0.00}{#1}}

\newcommand{\CommentTok}[1]{\textcolor[rgb]{0.38,0.63,0.69}{\textit{#1}}}

\newcommand{\DecValTok}[1]{\textcolor[rgb]{0.25,0.63,0.44}{#1}}

\newcommand{\FunctionTok}[1]{\textcolor[rgb]{0.02,0.16,0.49}{#1}}
\newcommand{\ImportTok}[1]{\textcolor[rgb]{0.00,0.50,0.00}{\textbf{#1}}}

\newcommand{\KeywordTok}[1]{\textcolor[rgb]{0.00,0.44,0.13}{\textbf{#1}}}
\newcommand{\NormalTok}[1]{#1}
\newcommand{\OperatorTok}[1]{\textcolor[rgb]{0.40,0.40,0.40}{#1}}
\newcommand{\OtherTok}[1]{\textcolor[rgb]{0.00,0.44,0.13}{#1}}

\newcommand{\StringTok}[1]{\textcolor[rgb]{0.25,0.44,0.63}{#1}}

\usepackage{microtype}

\ifx\paragraph\undefined\else
\let\oldparagraph\paragraph
\renewcommand{\paragraph}[1]{\oldparagraph{#1}\mbox{}}
\fi
\ifx\subparagraph\undefined\else
\let\oldsubparagraph\subparagraph
\renewcommand{\subparagraph}[1]{\oldsubparagraph{#1}\mbox{}}
\fi

\date{2023-05-22}

\begin{document}

\maketitle

\renewcommand{\shortauthors}{Grayson et al.}

\hypertarget{introduction}{%
\section{Introduction}\label{introduction}}

In practice, those seeking to reproduce a computational experiment often
need to manually look at the code to see how to build necessary
libraries, configure parameters, find data, and invoke the experiment;
it is not \emph{automatic}. Automatic reproducibility is a more
stringent goal, but working towards it would benefit the community. This
work discusses a machine-readable language for specifying how to execute
a computational experiment.

There is no existing standard place to put the ``main'' command which
executes an experiment. e.g., a Makefile which defines a rule for
executing the experiment alongside rules for compiling intermediate
pieces is not sufficient because there is no machine-readable way to
know which of the Make rules executes the experiment. Automatically
identifying the ``main'' command would be very useful for those seeking
to reproduce results from past experiments or reusing experiments to
address new use cases. For software engineering researchers, having a
standardized way to run many different codes at scale would open new
avenues for data mining research on reproducibility (c.f.,
\cite{collberg_repeatability_2016,trisovic_large-scale_2022,grayson_automatic_2023}).
We invite interested stakeholders to discuss this language at
\url{https://github.com/charmoniumQ/execution-description}.

Even with workflows, correctly invoking the experiment is still not
automatic. In a recent study, more than 70\% of workflows do not work
out-of-the-box \cite{grayson_automatic_2023}; for instance, they might
require the user to specify data or configure parameters for their
use-case. While flexibility is desirable, it should not preclude default
invocation in a standard location for testing purposes. For example, the
Snakemake workflow engine has a standard\footnote{See Snakemake Catalog
  rules for inclusion here:
  \url{https://snakemake.github.io/snakemake-workflow-catalog/?rules=true}}
for documenting the required arguments of its workflows, this standard
does not have a place to put an example invocation\footnote{See this
  discussion on GitHub:
  \url{https://github.com/snakemake-workflows/dna-seq-varlociraptor/pull/204\#issuecomment-1432876029}}.

\hypertarget{towards-a-standard-for-automatic-reproducibility}{%
\section{Towards a Standard for Automatic
Reproducibility}\label{towards-a-standard-for-automatic-reproducibility}}

There are many solutions for expressing how to run code, including bash
scripts, continuous integration scripts, workflows, and container
specifications. One can manually sleuth around how to run a handful of
experiments, but large-scale reproduction studies need to analyze
hundreds or thousands of codes
\cite{collberg_repeatability_2016,zhao_why_2012,grayson_automatic_2023}.
They each use different tools to invoke their experiment. Moreover, when
a code crashes in such a study, it is difficult to assess whether there
is a fault with the code or whether the study did not invoke the code as
intended. While we do not expect (or recommend) that the scientific
software community converge on a single solution for executing codes, we
see value in having a standard way of documenting how to run each code
that could hand off to the user's tool of choice.

One could implement such a specification using linked-data on the
semantic web. Defining the language in linked data lets one link to
existing data and reuse existing ontologies such as RO-crate
\cite{soiland-reyes_packaging_2022}, Dublin Core metadata terms
\cite{weibel_dublin_2000}, Description of a Project
\cite{wilder-james_description_2017}, nanopublications
\cite{groth_anatomy_2010}, Citation Typing Ontology
\cite{shotton_cito_2010}, and Document Components Ontology
\cite{constantin_document_2016}.

At the most basic level, the automatic reproducibility specification
should allow one to specify relevant commands and a string describing
their purpose (see \texttt{\#make} in Appendix \ref{code}). The strings
could be something like ``compile'', ``run'', or ``make-figures'', which
would be used the same way by multiple projects. However, the language
should go beyond fixed-strings.

The language should allow users to link commands directly to claims made
in publications (see \texttt{\#links-to-pub} in Appendix \ref{code}).
With such a specification, any person (or program) should be able to
execute the experiments which generate figures or claims in an
accompanying paper. For example, the CiTO vocabulary
\cite{shotton_cito_2010} can encode to how the result is used as
evidence in a specific publication.

The description can be even more granular than a publication. One could
use the DoCO vocabulary \cite{constantin_document_2016} to point to
specific figures, tables, or sentences within a document. Alternatively,
one could reference specific scientific claims using the Nanopublication
vocabulary \cite{groth_anatomy_2010} (see \texttt{\#links-to-fig},
\texttt{\#defines-nanopub}, and \texttt{\#links-to-nanopub} in Appendix
\ref{code}).

RO-crate \cite{soiland-reyes_wf4ever_2013} has terms for describing
dependencies between steps, which can be used to encode dependent steps
(see \texttt{\#make-data} and \texttt{\#plot-figures} in Appendix
\ref{code}). If the code requires a specific computational environment,
building that environment can be a prerequisite step. The purpose of
encoding dependencies is not to usurp the build-system or workflow
engine, which both already handle task dependencies; if the experiment
already uses a workflow, then the specification should invoke that. The
purpose of task dependencies in the specification is for projects which
do not use a workflow engine, or a task that installs the desired
workflow engine.

Such a specification could also set bounds on the experiment's
parameters, such as the range of valid values or a list of toggleable
parameters (see \texttt{\#example-of-parameters} in Appendix
\ref{code}). This parameter metadata would enable downstream automated
experiments like parameter-space search studies, multi-fidelity
uncertainty quantification, and outcome-preserving input minimization.

\hypertarget{getting-adoption}{%
\section{Getting Adoption}\label{getting-adoption}}

The most useful part of the specification would need \emph{some} human
input to create, which is specifying what the tasks do. However, we can
reduce the manual effort needed to write the specification.

Workflow engines could assist in generating this, since they know all
the computational steps, inputs, outputs, and parameters. Then it could
prompt the user with high-level questions (e.g., ``What publication is
this part of''?) and generate the appropriate specification to invoke
themselves.

If the experiment does not use a workflow engine, but someone who can
run the experiment is available, an interactive shell session can
capture and write the specification. The user would invoke a shell that
records every command, its exit status, its read-files, and its
write-files using syscall interposition. The user would run their code
as usual, and after finishing, the shell would assemble the necessary
computational steps and prompt the user for high-level questions.

As a last resort, if one finds a publication linking to a specific
repository, one can try to guess the main command. This approach is the
current state-of-the-art for large-scale reproduction studies, except a
standardized language would allow some large-scale reproduction studies
to inform future large-scale reproduction studies on what they did to
execute this repository. Computational scientists at least had an
opportunity to influence how to invoke their code in large-scale
reproduction studies. The lack of opportunity for input was a frequent
response of scientists to Collberg and Proebsting\footnote{The authors
  of publications whose labels are BarowyCBM12, BarthePB12,
  HolewinskiRRFPRS12, and others responded to Collberg and Proebsting
  (paraphrasing), ``it would have worked; you just didn't invoke the
  right commands.'' according to
  \url{http://reproducibility.cs.arizona.edu/v2/index.html}.}.

Computational scientists could benefit from creating these automated
reproducibility specifications because large-scale reproduction studies
like Collberg and Proebsting \cite{collberg_repeatability_2016}, Zhao et
al.~\cite{zhao_why_2012}, and others serve as free testing and
reproduction of their results.

Ideally, the reproduction specification would be placed in the same
location as the computational experiment, often a GitHub repository, so
developers can maintain it alongside the code. In cases where the
authors of the GitHub repository are not cooperative, one can instead
put reproduction specifications in a repository that holds reproduction
specifications written by the community, a ``reproducibility library''.
Users seeking to reproduce a repository would invoke a tool that looks
for an automatic reproducibility specification in the source code
repository, in a list of reproducibility libraries, and if none is
found, falls back on heuristic to guess how to reproduce the experiment.
The heuristic might have cases such as, ``if a Make file exists, run
\texttt{make\ all}''. If the fallback succeeds, the tool can upload all
its steps to a reproducibility library.

Meanwhile, conferences and publishers could promote such standard
specifications as part of reproducibility requirements for publishing.
Currently, to get an artifact evaluation badge, computational scientists
would have to write a natural language description of the software
environment, what the commands are, how to run them, and where the data
end up; meanwhile, an artifact evaluator has to read, interpret, and
execute their description by hand. An execution description could make
this automatic; if an execution description exists, the artifact
evaluator uses an executor which understands the language and runs all
of the commands that reference the manuscript in their \texttt{purpose}
tag.

\hypertarget{conclusion}{%
\section{Conclusion}\label{conclusion}}

Developing common standards for specifying how to run computational
experiments would benefit the scientific community. It presents a
compromise where different teams can implement their codes however they
see fit while enabling others to run them easily. This specification
would lead to greater productivity in the (re)use of scientific
experiments, empower developers to build tools that leverage those
common specifications, and enable software engineering researchers to
study reproducibility at scale.

\eject

\bibliographystyle{ACM-Reference-Format}
\bibliography{manual,common/sams-zotero-export}

\appendix

\hypertarget{listing-of-an-example-reproducibility-specification}{%
\section{Listing of an Example Reproducibility
Specification}\label{listing-of-an-example-reproducibility-specification}}

\label{code}

The following language sample is not the final proposal for the complete
vocabulary; the peer-review process is not ideal to iterate on technical
details. Instead, we invite technical contributions at the repository,
\url{https://github.com/charmoniumQ/execution-description}. The point of
this article is to argue that the community should spend effort
developing this vocabulary.

\small

\begin{Shaded}
\begin{Highlighting}[]
\FunctionTok{\textless{}?xml}\OtherTok{ version=}\StringTok{"1.0"}\OtherTok{ encoding=}\StringTok{"utf{-}8"}\FunctionTok{?\textgreater{}}
\CommentTok{\textless{}!{-}{-}}
\CommentTok{RDF can be serialized as XML, JSON, or triples; backend RDF parsers don\textquotesingle{}t care.}
\CommentTok{We chose XML because it might be more familiar to readers.}
\CommentTok{{-}{-}\textgreater{}}

\CommentTok{\textless{}!{-}{-}}
\CommentTok{The following tag imports several other vocabularies behind a namespace.}
\CommentTok{E.g., \textasciigrave{}rdf:type\textasciigrave{} refers to \textasciigrave{}type\textasciigrave{} in the \textasciigrave{}rdf\textasciigrave{} namespace, which resolves to:}
\CommentTok{http://www.w3.org/1999/02/22{-}rdf{-}syntax{-}ns\#rdftype}
\CommentTok{Elements with no namespace are resolved within the default namespace,}
\CommentTok{which is our proposed execution{-}description vocabulary, http://example.org/execution{-}description/1.0.}
\CommentTok{{-}{-}\textgreater{}}

\NormalTok{\textless{}}\KeywordTok{rdf:RDF}\OtherTok{ xmlns:rdf=}\StringTok{"http://www.w3.org/1999/02/22{-}rdf{-}syntax{-}ns\#"}
\OtherTok{         xmlns:rdfs=}\StringTok{"http://www.w3.org/2000/01/rdf{-}schema\#"}
\OtherTok{         xmlns:dc=}\StringTok{"http://purl.org/dc/elements/1.1/"}
\OtherTok{         xmlns:wikibase=}\StringTok{"http://wikiba.se/ontology\#"}
\OtherTok{         xmlns:cito=}\StringTok{"http://purl.org/spar/cito"}
\OtherTok{         xmlns:doco=}\StringTok{"http://purl.org/spar/doco/2015{-}07{-}03"}
\OtherTok{         xmlns:prov=}\StringTok{"http://www.w3.org/TR/2013/PR{-}prov{-}o{-}20130312/"}
\OtherTok{         xmlns:wfdesc=}\StringTok{"http://purl.org/wf4ever/wfdesc\#"}
\OtherTok{         xml:lang=}\StringTok{"en"}
\NormalTok{         \textgreater{}}

  \CommentTok{\textless{}!{-}{-}}
\CommentTok{  Here, we list some relevant commands, and how they relate to the artifact.}
\CommentTok{  {-}{-}\textgreater{}}
\NormalTok{  \textless{}}\KeywordTok{process}\OtherTok{ rdf:about=}\StringTok{"\#make"}\NormalTok{\textgreater{}}
    \CommentTok{\textless{}!{-}{-} The following would get run by the UNIX shell. {-}{-}\textgreater{}}
\NormalTok{    \textless{}}\KeywordTok{command}\NormalTok{\textgreater{}make libs\textless{}/}\KeywordTok{command}\NormalTok{\textgreater{}}
    \CommentTok{\textless{}!{-}{-} Here is a string representing the purpose. {-}{-}\textgreater{}}
\NormalTok{    \textless{}}\KeywordTok{purpose}\NormalTok{\textgreater{}compiles libraries\textless{}/}\KeywordTok{purpose}\NormalTok{\textgreater{}}
\NormalTok{  \textless{}/}\KeywordTok{process}\NormalTok{\textgreater{}}

  \CommentTok{\textless{}!{-}{-}}
\CommentTok{  Here, we make a process that depends on a previous process using wfdesc.}
\CommentTok{  {-}{-}\textgreater{}}
\NormalTok{  \textless{}}\KeywordTok{process}\OtherTok{ rdf:about=}\StringTok{"\#make{-}data"}\NormalTok{\textgreater{}}
\NormalTok{    \textless{}}\KeywordTok{command}\NormalTok{\textgreater{}python3 make\_data.py\textless{}/}\KeywordTok{command}\NormalTok{\textgreater{}}
\NormalTok{    \textless{}}\KeywordTok{purpose}\NormalTok{\textgreater{}makes data\textless{}/}\KeywordTok{purpose}\NormalTok{\textgreater{}}
\NormalTok{  \textless{}/}\KeywordTok{process}\NormalTok{\textgreater{}}
\NormalTok{  \textless{}}\KeywordTok{process}\OtherTok{ rdf:about=}\StringTok{"\#plot{-}figures"}\NormalTok{\textgreater{}}
\NormalTok{    \textless{}}\KeywordTok{command}\NormalTok{\textgreater{}python3 figures.py\textless{}/}\KeywordTok{command}\NormalTok{\textgreater{}}
\NormalTok{    \textless{}}\KeywordTok{purpose}\NormalTok{\textgreater{}plot figures\textless{}/}\KeywordTok{purpose}\NormalTok{\textgreater{}}
\NormalTok{    \textless{}}\KeywordTok{dependsOn}\OtherTok{ rdf:resource=}\StringTok{"\#make{-}data"}\NormalTok{ /\textgreater{}}
    \CommentTok{\textless{}!{-}{-}}
\CommentTok{    The \# is not a typo; the rdf:about becomes a URL fragment in the current document.}
\CommentTok{    This means one can access a computational step in another document here,}
\CommentTok{    like "https://example.com/software{-}experiment{-}23\#make{-}data".}
\CommentTok{    {-}{-}\textgreater{}}
\NormalTok{  \textless{}/}\KeywordTok{process}\NormalTok{\textgreater{}}
  \CommentTok{\textless{}!{-}{-} Users may choose the more complex wfdesc vocabulary if they wish. {-}{-}\textgreater{}}

  \CommentTok{\textless{}!{-}{-}}
\CommentTok{  Links to a publication.}
\CommentTok{  The publisher may or may not host a linked{-}data description of the documenta at this URL.}
\CommentTok{  The purpose of the URL is to unambiguously name the document.}
\CommentTok{  We need the rdf:Description to reference an external resource.}
\CommentTok{  {-}{-}\textgreater{}}
\NormalTok{  \textless{}}\KeywordTok{process}\OtherTok{ rdf:about=}\StringTok{"links{-}to{-}pub"}\NormalTok{\textgreater{}}
\NormalTok{    \textless{}}\KeywordTok{command}\NormalTok{\textgreater{}make all\textless{}/}\KeywordTok{command}\NormalTok{\textgreater{}}
\NormalTok{    \textless{}}\KeywordTok{purpose}\NormalTok{\textgreater{}}
\NormalTok{      \textless{}}\KeywordTok{rdf:Description}\NormalTok{\textgreater{}}
\NormalTok{        \textless{}}\KeywordTok{cito:isCitedAsEvidenceBy}\OtherTok{ rdf:resource=}\StringTok{"https://doi.org/10.1234/123456789"}\NormalTok{ /\textgreater{}}
\NormalTok{      \textless{}/}\KeywordTok{rdf:Description}\NormalTok{\textgreater{}}
\NormalTok{    \textless{}/}\KeywordTok{purpose}\NormalTok{\textgreater{}}

  \CommentTok{\textless{}!{-}{-} Links to a specific figure within a publication {-}{-}\textgreater{}}
\NormalTok{  \textless{}}\KeywordTok{process}\OtherTok{ rdf:about=}\StringTok{"links{-}to{-}fig"}\NormalTok{\textgreater{}}
\NormalTok{    \textless{}}\KeywordTok{command}\NormalTok{\textgreater{}make all\textless{}/}\KeywordTok{command}\NormalTok{\textgreater{}}
\NormalTok{    \textless{}}\KeywordTok{purpose}\NormalTok{\textgreater{}}
\NormalTok{      \textless{}}\KeywordTok{prov:generated}\NormalTok{\textgreater{}}
\NormalTok{        \textless{}}\KeywordTok{doco:figure}\NormalTok{\textgreater{}}
\NormalTok{          \textless{}}\KeywordTok{rdf:Description}\NormalTok{\textgreater{}}
\NormalTok{            \textless{}}\KeywordTok{dc:title}\NormalTok{\textgreater{}Figure 2b\textless{}/}\KeywordTok{dc:title}\NormalTok{\textgreater{}}
\NormalTok{            \textless{}}\KeywordTok{dc:isPartOf}\OtherTok{ rdf:resource=}\StringTok{"https://doi.org/10.1234/123456789"}\NormalTok{ /\textgreater{}}
\NormalTok{          \textless{}/}\KeywordTok{rdf:Description}\NormalTok{\textgreater{}}
\NormalTok{        \textless{}/}\KeywordTok{doco:figure}\NormalTok{\textgreater{}}
\NormalTok{      \textless{}/}\KeywordTok{prov:generated}\NormalTok{\textgreater{}}
\NormalTok{    \textless{}/}\KeywordTok{purpose}\NormalTok{\textgreater{}}

  \CommentTok{\textless{}!{-}{-}}
\CommentTok{  Describes an abstract nanopublication claim that this experiment supports.}
\CommentTok{  This one will say: "this experiment supports the claim that malaria is spread by mosquitoes"}
\CommentTok{  {-}{-}\textgreater{}}
\NormalTok{  \textless{}}\KeywordTok{process}\OtherTok{ rdf:about=}\StringTok{"defines{-}nanopub"}\NormalTok{\textgreater{}}
\NormalTok{    \textless{}}\KeywordTok{command}\NormalTok{\textgreater{}make all\textless{}/}\KeywordTok{command}\NormalTok{\textgreater{}}
\NormalTok{    \textless{}}\KeywordTok{purpose}\NormalTok{\textgreater{}}
\NormalTok{      \textless{}}\KeywordTok{cito:supports}\NormalTok{\textgreater{}}
        \CommentTok{\textless{}!{-}{-}}
\CommentTok{        We will use Wikidata here.}
\CommentTok{        They have catalogued many real{-}world objects and concepts as linked{-}data objects.}
\CommentTok{        {-}{-}\textgreater{}}
\NormalTok{        \textless{}}\KeywordTok{wikibase:Statement}\NormalTok{\textgreater{}}
\NormalTok{          \textless{}}\KeywordTok{rdf:Description}\NormalTok{\textgreater{}}
            \CommentTok{\textless{}!{-}{-} Q12156 refers to malaria {-}{-}\textgreater{}}
\NormalTok{            \textless{}}\KeywordTok{subject}\OtherTok{ rdf:resource=}\StringTok{"https://www.wikidata.org/entity/Q12156"}\NormalTok{ /\textgreater{}}
            \CommentTok{\textless{}!{-}{-} P1060 refers to disease transmission process (read: "is transmitted by") {-}{-}\textgreater{}}
\NormalTok{            \textless{}}\KeywordTok{predicate}\OtherTok{ rdf:resource=}\StringTok{"http://www.wikidata.org/prop/P1060"}\NormalTok{ /\textgreater{}}
            \CommentTok{\textless{}!{-}{-} Q15304532 refers to mosquitoes {-}{-}\textgreater{}}
\NormalTok{            \textless{}}\KeywordTok{object}\OtherTok{ rdf:resource=}\StringTok{"https://www.wikidata.org/entity/Q15304532"}\NormalTok{ /\textgreater{}}
\NormalTok{          \textless{}/}\KeywordTok{rdf:Description}\NormalTok{\textgreater{}}
\NormalTok{        \textless{}/}\KeywordTok{wikibase:Statement}\NormalTok{\textgreater{}}
\NormalTok{      \textless{}/}\KeywordTok{cito:supports}\NormalTok{\textgreater{}}
\NormalTok{    \textless{}/}\KeywordTok{purpose}\NormalTok{\textgreater{}}

    \CommentTok{\textless{}!{-}{-}}
\CommentTok{    Alternatively, the nanopublication claim will live somewhere else.}
\CommentTok{    Linked data lets us seamlessly reference other documents.}
\CommentTok{    {-}{-}\textgreater{}}
\NormalTok{    \textless{}}\KeywordTok{purpose}\OtherTok{ rdf:about=}\StringTok{"links{-}to{-}nanopub"}\NormalTok{\textgreater{}}
\NormalTok{      \textless{}}\KeywordTok{rdf:Description}\NormalTok{\textgreater{}}
\NormalTok{        \textless{}}\KeywordTok{cito:supports}\OtherTok{ rdf:resource=}\StringTok{"https://example.com/article24\#claim31"}\NormalTok{ /\textgreater{}}
\NormalTok{      \textless{}/}\KeywordTok{rdf:Description}\NormalTok{\textgreater{}}
\NormalTok{    \textless{}/}\KeywordTok{purpose}\NormalTok{\textgreater{}}
\NormalTok{  \textless{}/}\KeywordTok{process}\NormalTok{\textgreater{}}

  \CommentTok{\textless{}!{-}{-} Here, we add parameters to the command {-}{-}\textgreater{}}
\NormalTok{  \textless{}}\KeywordTok{process}\OtherTok{ rdf:label=}\StringTok{"example{-}of{-}parameters"}\NormalTok{\textgreater{}}
    \CommentTok{\textless{}!{-}{-} These might be template filled like so: {-}{-}\textgreater{}}
\NormalTok{    \textless{}}\KeywordTok{command}\NormalTok{\textgreater{}./generate $\{max\_resolution\} $\{rounds\}\textless{}/}\KeywordTok{command}\NormalTok{\textgreater{}}
\NormalTok{    \textless{}}\KeywordTok{wfdesc:Parameter}\OtherTok{ rdfs:label=}\StringTok{"max\_resolution"}\NormalTok{ /\textgreater{}}
\NormalTok{  \textless{}/}\KeywordTok{process}\NormalTok{\textgreater{}}

\NormalTok{\textless{}/}\KeywordTok{rdf:RDF}\NormalTok{\textgreater{}}
\end{Highlighting}
\end{Shaded}

\normalsize

The above RDF/XML can be validated with Python and rdflib:

\begin{Shaded}
\begin{Highlighting}[]
\OperatorTok{\textgreater{}\textgreater{}\textgreater{}} \ImportTok{import}\NormalTok{ rdflib}
\OperatorTok{\textgreater{}\textgreater{}\textgreater{}}\NormalTok{ g }\OperatorTok{=}\NormalTok{ rdflib.Graph().parse(}\StringTok{"test.xml"}\NormalTok{)}
\OperatorTok{\textgreater{}\textgreater{}\textgreater{}} \CommentTok{\# Now we can iterate over the triples contained in this RDF graph}
\OperatorTok{\textgreater{}\textgreater{}\textgreater{}} \CommentTok{\# Note that "anonymous nodes" will appear as rdflib.term.BNode(\textquotesingle{}...\textquotesingle{})}
\OperatorTok{\textgreater{}\textgreater{}\textgreater{}} \BuiltInTok{list}\NormalTok{(g)[:}\DecValTok{5}\NormalTok{]}
\NormalTok{[(rdflib.term.BNode(}\StringTok{\textquotesingle{}N979c272652c948f48598caa65eaf02da\textquotesingle{}}\NormalTok{),}
\NormalTok{  rdflib.term.URIRef(}\StringTok{\textquotesingle{}http://www.w3.org/1999/02/22{-}rdf{-}syntax{-}ns\#type\textquotesingle{}}\NormalTok{),}
\NormalTok{  rdflib.term.URIRef(}\StringTok{\textquotesingle{}http://www.w3.org/TR/2013/PR{-}prov{-}o{-}20130312/generated\textquotesingle{}}\NormalTok{)),}
\NormalTok{ (rdflib.term.URIRef(}\StringTok{\textquotesingle{}file:///home/sam/box/execution{-}description/se4rs/test.xml\#plot{-}figures\textquotesingle{}}\NormalTok{),}
\NormalTok{  rdflib.term.URIRef(}\StringTok{\textquotesingle{}file:///.../purpose\textquotesingle{}}\NormalTok{),}
\NormalTok{  rdflib.term.Literal(}\StringTok{\textquotesingle{}plot figures\textquotesingle{}}\NormalTok{, lang}\OperatorTok{=}\StringTok{\textquotesingle{}en\textquotesingle{}}\NormalTok{)),}
\NormalTok{ (rdflib.term.BNode(}\StringTok{\textquotesingle{}N979c272652c948f48598caa65eaf02da\textquotesingle{}}\NormalTok{),}
\NormalTok{  rdflib.term.URIRef(}\StringTok{\textquotesingle{}http://purl.org/spar/doco/2015{-}07{-}03figure\textquotesingle{}}\NormalTok{),}
\NormalTok{  rdflib.term.BNode(}\StringTok{\textquotesingle{}Ned5bd1d9a83b48bfa0798f2f1e296db7\textquotesingle{}}\NormalTok{)),}
\NormalTok{ (rdflib.term.BNode(}\StringTok{\textquotesingle{}Nc4f1068252194a4d90b91a02f3860cf7\textquotesingle{}}\NormalTok{),}
\NormalTok{  rdflib.term.URIRef(}\StringTok{\textquotesingle{}http://wikiba.se/ontology\#Statement\textquotesingle{}}\NormalTok{),}
\NormalTok{  rdflib.term.BNode(}\StringTok{\textquotesingle{}Nce17a7a5920846788169b713dd655c97\textquotesingle{}}\NormalTok{)),}
\NormalTok{ (rdflib.term.BNode(}\StringTok{\textquotesingle{}N889f577571ab4c67bc063a0d032eb5cf\textquotesingle{}}\NormalTok{),}
\NormalTok{  rdflib.term.URIRef(}\StringTok{\textquotesingle{}file:///.../purpose\textquotesingle{}}\NormalTok{),}
\NormalTok{  rdflib.term.BNode(}\StringTok{\textquotesingle{}Nc4f1068252194a4d90b91a02f3860cf7\textquotesingle{}}\NormalTok{))]}
\end{Highlighting}
\end{Shaded}

\end{document}